\documentclass[12pt]{aastex}

\usepackage{ifthen}

\usepackage{natbib, latexsym}
\usepackage{graphics, lscape}
\usepackage{rotating}
\usepackage{txfonts}

\def\mum{${\rm \mu m}$}
\def\arcsec{\hbox{$^{\prime\prime}$}}

\def\HII{HII}

\shorttitle{Magic PAHs}
\shortauthors{Peeters et al.}
\begin{document}
\title {The 15-20 \mum\, emission in the reflection nebula NGC\,2023}
\author{Els Peeters\altaffilmark{1,2}, Alexander G.G.M. Tielens\altaffilmark{3}, Louis J. Allamandola\altaffilmark{4}, Mark G. Wolfire\altaffilmark{5}}
\altaffiltext{1}{Department of Physics and Astronomy, University of Western Ontario, London, ON N6A 3K7, Canada;
epeeters@uwo.ca}
\altaffiltext{2}{SETI Institute, 189 Bernardo Avenue, Suite 100, Mountain View, CA 94043, USA}
\altaffiltext{3}{Leiden Observatory, PO Box 9513, 2300 RA Leiden, The Netherlands; tielens@strw.leidenuniv.nl}
\altaffiltext{4}{NASA-Ames Research Center, Space Science Division, Mail Stop 245-6, Moffett Field, CA 94035, USA; Louis.J.Allamandola@nasa.gov}
\altaffiltext{5}{Astronomy Department, University of Maryland, College Park, MD 20742, USA; mwolfire@astro.umd.edu}


\begin{abstract}

We present 15-20 \mum\, spectral maps towards the reflection nebula NGC\,2023 obtained with the Infrared Spectrograph in short-wavelength, high-resolution mode on board the Spitzer Space Telescope. These spectra reveal emission from PAHs, C$_{60}$, and H$_2$ superposed on a dust continuum. These emission components exhibit distinct spatial distributions: with increasing distance from the illuminating star, we observe the PAH emission followed by the dust continuum emission and the H$_2$ emission. The C$_{60}$ emission is located closest to the illuminating star in the south while in the north, it seems to be associated with the H/H$_2$ transition. Emission from PAHs and PAH-related species produce features at 15.8, 16.4, 17.4, and 17.8 \mum\, and the 15-18 \mum\, plateau. These different PAH features show distinct spatial distributions. The 15.8 \mum\, band and 15-18 \mum\, plateau correlate with the 11.2 \mum\, PAH band and with each other, and are attributed to large, neutral PAHs. Conversely, the 16.4 \mum\, feature correlates with the 12.7 \mum\, PAH band, suggesting that both arise from species that are favored by the same conditions that favor PAH cations. The PAH contribution to the 17.4 \mum\, band is displaced towards the illuminating star with respect to the 11.2 and 12.7 \mum\, emission and is assigned to doubly ionized PAHs and/or a subset of cationic PAHs. The spatial distribution of the 17.8 \mum\, band suggests it arises from both neutral and cationic PAHs. In contrast to their intensities, the profiles of the PAH bands and the 15-18 \mum\, plateau do not vary spatially. Consequently, we conclude that the carrier of the 15-18 \mum\, plateau is distinct from that of the PAH bands. 
 
\end{abstract}

\section{Introduction}
\label{intro}

Strong emission bands at 3.3, 6.2, 7.7 and 11.3 \mum\,- the so-called unidentified infrared (UIR) bands - dominate the mid-IR spectra of almost all objects, including reflection nebulae, planetary nebulae, the interstellar medium and \HII\, regions. These bands are generally attributed to IR fluorescence of a family of Polycyclic Aromatic Hydrocarbon molecules (PAHs) pumped by the UV radiation field. A key result in the observational studies of PAHs is that their mid-IR bands show clear variations in peak positions, shapes and (relative) intensities, not only between sources, but also spatially within extended sources \citep[e.g.][]{Hony:oops:01, Peeters:prof6:02, Brandl:06, SmithJD:07, Galliano:08}. The variability in intensity ratio of the main PAH bands has been interpreted in terms of the charge state and excitation levels of the PAHs \citep[e.g.][]{Allamandola:modelobs:99, Galliano:08} while the variation in the profiles of the main PAH bands \citep[classified as class A, B and C profiles;][]{Peeters:prof6:02, vanDiedenhoven:chvscc:03} is attributed to the environment \citep[CSM versus ISM;][]{Peeters:prof6:02, VanKerckhoven:phd:02, Boersma:08} and the degree of UV processing \citep{Sloan:07, Tielens:08, Boersma:08}.

  In addition to the well-studied main PAH bands, a plethora of weaker PAH bands are present, for example in the 15-20 \mum\, region. PAH emission in this region was first observed with the Infrared Space Observatory (ISO): new features were reported at 15.8, 16.4, 17.4 and 17.9 \mum\, with the 16.4 and then 17.4 \mum\, PAH bands being the most prominent \citep{Beintema:pahs:96, Moutou:16.4:00, Sturm:swsgal:00, VanKerckhoven:plat:00}. In addition, \citet{VanKerckhoven:plat:00} presented evidence for a variable, broad plateau from 15 to 20 \mum\, which seems to be present solely in \HII\, regions \citep{VanKerckhoven:plat:00, Peeters:plat:04, Peeters:spitzer:04}. Based upon observations with the Spitzer Space Telescope, additional broad components are reported at 16.6, 17.0 and 17.2 \mum\, \citep{SmithJD:04, SmithJD:07, Sellgren:07}. But it is really the high sensitivity of Spitzer that revealed the omnipresence of these bands within the Milky Way and in other galaxies \citep[e.g.][]{Werner:04, SmithJD:04, Brandl:04, Sellgren:07, SmithJD:07, Tappe:06}, allowing - for the first time -  systematic investigations of these weaker bands.  

The 15-20 \mum\, emission is attributed to C-C-C bending vibrations \citep[e.g.][]{ATB, Moutou:firempahs:96, VanKerckhoven:plat:00, Peeters:plat:04, Mattioda:09, Boersma:10, Ricca:10}. As a consequence, the bands in the 15-20 \mum\, region are expected to reveal more about the overall molecular structure of the carriers than do the major bands below 15 \mum. An intriguing result of the Spitzer observations is the remarkable similarity (to first order) of the 15-20 \mum\, PAH emission spectra from regions spanning a large range of physical conditions. This suggests that the astronomical PAH population may be dominated by a handful of stable, molecular structures (cf. survival of the fittest) rather than comprised of a large number of PAHs with widely varying structures \citep{Boersma:10}. Hence, a systematic study of a large sample of spectra showing these weaker PAH bands will provide a complementary view on the characteristics of the emitting PAH population. 

This paper reports such a study.  Here we analyze Spitzer-IRS spectral maps of NGC\,2023 in the 15-20 \mum\, region. In Section \ref{source},  we describe the reflection nebula NGC\,2023, while the observations and data reduction are discussed in Section \ref{data}. The data analysis is presented in Section \ref{analysis} and discussed in Section \ref{discussion}. We end with a short summary in Section \ref{conclusion}.

\begin{figure}[t!]
    \centering
\resizebox{14cm}{!}{%
  \includegraphics{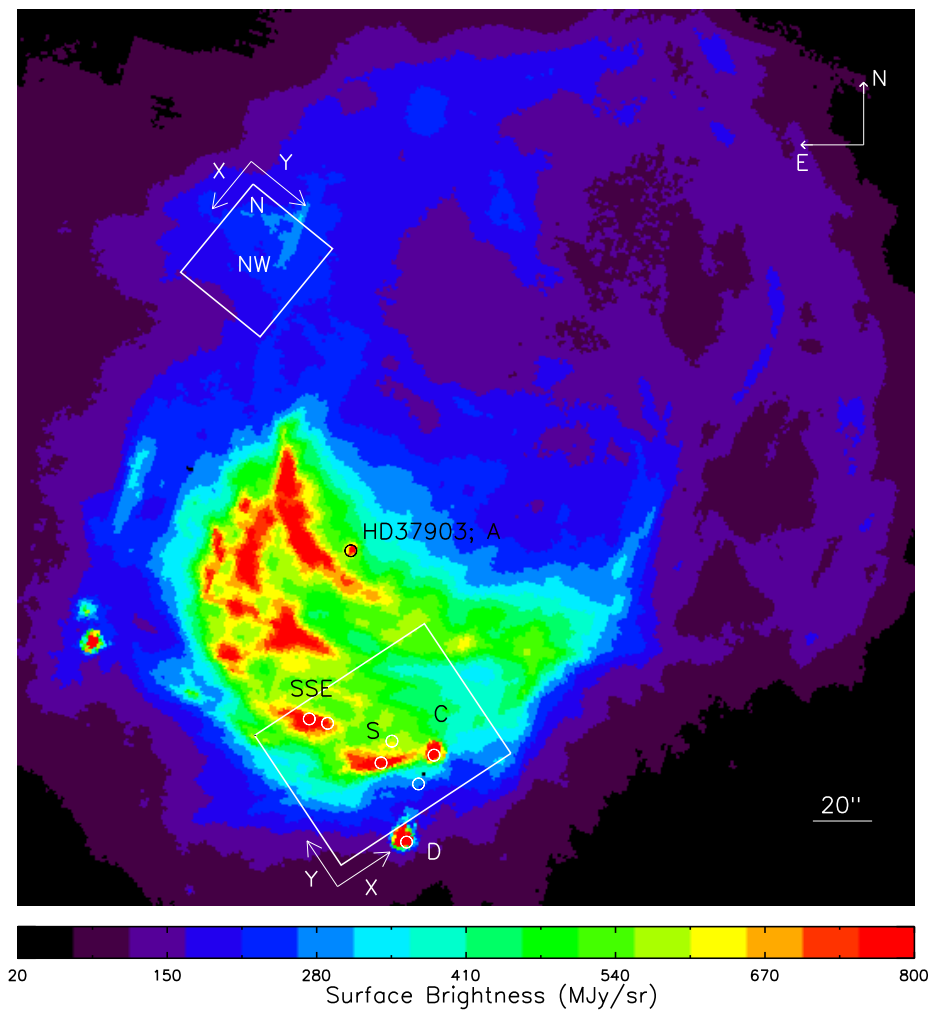}}
\caption{The IRAC [8.0] image of NGC\,2023 with the SH FOV shown (white) for both the north and south positions studied here. The star HD37903 is indicated by a black circle, sources A, C and D are from \citet{Sellgren:83}, and the white circles indicate 2MASS point sources located inside the SH apertures. S refers to the south ridge, SSE to the south-southeastern ridge, N to the north ridge, and NW to the northwestern ridge. Maps shown in this paper use the orientation denoted by the vectors (X, Y).}
\label{fov}
\end{figure}

\section{The reflection nebula NGC\,2023}
\label{source}
NGC\,2023, one of the best studied reflection nebulae, is also among the brightest in the IR \citep[e.g.][]{Knapp:75, Harvey:80, Sellgren:84, Witt:84, Sellgren:85, Gatley:87, Witt:89, Howe:91, Jaffe:94, Sellgren:96, Steiman:97, Field:98, Martini:99, McCartney:99, Mookerjea:00, Takami:00, Wyrowski:00, Burgh:02, Werner:04, Sellgren:07, Sellgren:10}. Illuminated by the B1.5 V star HD\,37903, it is at a distance of 350 pc \citep{Mookerjea:09, Sheffer:11}. Images reveal a network of narrow filaments and hot spots (see e.g. Fig.\,\ref{fov}); the width of the H$_2$ filaments is $\sim$0.25-0.5\arcsec \citep{Rouan:97}.  The amount of morphological detail in these images is particularly striking. While the overall morphology is similar at all wavelengths, there are significant variations in location and contrast between filaments and background in the different tracers. The limb-brightened shells so prominent in the various maps of this reflection nebula provide a unique opportunity to study the layered spatial distribution of Photo-Dissociation Regions (PDRs).  Detailed analysis of the emission in sub-mm, CO rotational, FIR fine structure, radio recombination lines and NIR H$_2$ lines \citep{Jaffe:90, Burton:90, Steiman:97, Wyrowski:97, Wyrowski:00} in a few selected locations using PDR models \citep[e.g.][]{Tielens:PDR:85} imply that a FUV radiation field of 10$^4$ times the average interstellar value is incident on a clumpy molecular cloud, with the bulk of the gas having densities varying from 10$^3$ to 10$^5$ cm$^{-3}$, depending on location.

\begin{table}[t!]
\small
\caption{\label{log} Observation log}
\begin{center}
\begin{tabular}{c|c|c}
 & north position  & south position \\
 \hline
 \hline
 & &  \\[-5pt]

PID & 50511  & 20097\\
AOR &  26337024 &   14033920 \\
cycles x ramp time& 2x30s  & 1x30s \\
pointings $\parallel$ & 7 & 12\\
step size $\parallel$ & 5\arcsec &   5.65\arcsec\\
pointings $\bot$ & 15 &  12\\
step size $\bot$ & 2.3\arcsec & 4.7\arcsec\\
background$^{a}$ & 5:42:1.00, -2:6:54.5 &  \\
\hline
 \multicolumn{2}{c}{} \\[-5pt]

\multicolumn{3}{c}{$^a$ $\alpha, \delta$ (J2000); units of $\alpha$ are hours, minutes, and seconds, and}\\
\multicolumn{3}{c}{units of $\delta$ are degrees, arc minutes, and arc seconds.}
\end{tabular} 
\end{center}
\end{table}

\section{The data}
\label{data}

\subsection{Observations}
\label{obs}

The observations were taken with the Infrared Spectrograph \citep[IRS,][]{houck04} on board the Spitzer Space Telescope \citep{werner04} and were part of the Open Time Observations PIDs 20097 and 50511.

 We obtained spectral maps for two positions in the reflection nebula NGC\,2023 (see Fig.\,\ref{fov}): towards the dense shell 78\arcsec\, south of the exciting star HD 37903 corresponding to the H$_2$ emission peak, and towards a region to the north (+33\arcsec, +105\arcsec) of the exciting star \citep{Burton:98}. The north position is characterized by a much lower density of $\sim$ 10$^{4}$ cm$^{-3}$ than the density at the south position, which is $>$ 10$^{5}$ cm$^{-3}$ \citep{Burton:98, Sheffer:11}. Likewise, the north position has a slightly lower UV radiation field between 6 and 13.6 eV, i.e. $\sim$ 500 G$_0$\footnote{G$_0$ is the integrated 6 to 13.6 eV radiation flux in
units of the Habing field = 1.6x10$^{-3}$ ergs/cm$^2$/s.}, as compared to $\sim$ 10$^4$ G$_0$ at the southern H$_2$ peak \citep{Burton:98, Sheffer:11}.

The spectral maps are made with the short-high (SH) mode. The SH mode covers a wavelength range from 10 to 20 \mum\, at a resolution of $\sim$ 600 and has a pixel size of 2.3\arcsec. We obtained off-source background observations in SH for the north position. Table \ref{log} gives a detailed overview of the observations. We also obtained spectral maps made with the short-low (SL) mode; the combined SL \& SH data set will be presented in Peeters et al. (2012, in prep.).

\subsection{Reduction}
The raw data were processed with the S18.7 pipeline version by the Spitzer Science Center. The resulting bcd-products are further processed using {\it cubism} \citep{cubism} and {\it irsclean} available from the SSC website\footnote{http://ssc.spitzer.caltech.edu}.

The SH off-source background observation exhibits a rising dust continuum and an 11.2 \mum\, PAH band. Since we do not have off-source background observations for the south position, we investigated the importance of the background flux and, in particular, of the background PAH emission with respect to the on-source (PAH) flux. The 11.2 PAH flux in the SH background for the northern position is 8.5 10$^{-8}$ W/m$^{2}$/sr. This amounts to 3-11\% of the on-source 11.2 PAH flux across the north map and to $<$1-10\% of the on-source 11.2 PAH flux across the south map excluding source D from \citet{Sellgren:83}. Source D is located on the edge of the south map and has a background contribution to the PAH flux of up to 45\%. These results are consistent with the IRAC [8 \mum] image which is dominated by PAH emission and indicates a background flux of $\sim$ 6\% the {\it minimum} on-source flux. Hence, aside from source D, we conclude that the background has only a small contribution to the on-source PAH flux and therefore we did not apply a background subtraction. For the remaining of this paper, we  therefor excluded source D from the PAH analysis.

Since no background subtraction was applied, we used the tool {\it irsclean} to correct for the presence of rogue pixels. These rogue pixels are particularly problematic for the high-resolution module SH. In addition, we applied {\it cubism}'s automatic bad pixel generation with $\sigma_{TRIM} = 7$ and Minbad-fraction = 0.50 and 0.75 for the global bad pixels and record bad pixels respectively. Remaining bad pixels were subsequently removed manually. 

Spectra were extracted from the spectral maps by moving, in one pixel steps, a spectral aperture of 2x2 pixels in both directions of the maps. This results in overlapping extraction apertures. No jumps in flux level were seen between the different orders within the SH module (SH11 to SH20) and hence no scaling was necessary.

\subsection{The spectra}
\label{sp}
\begin{figure}[t!]
    \centering
\resizebox{0.35\textwidth}{!}{%
  \includegraphics{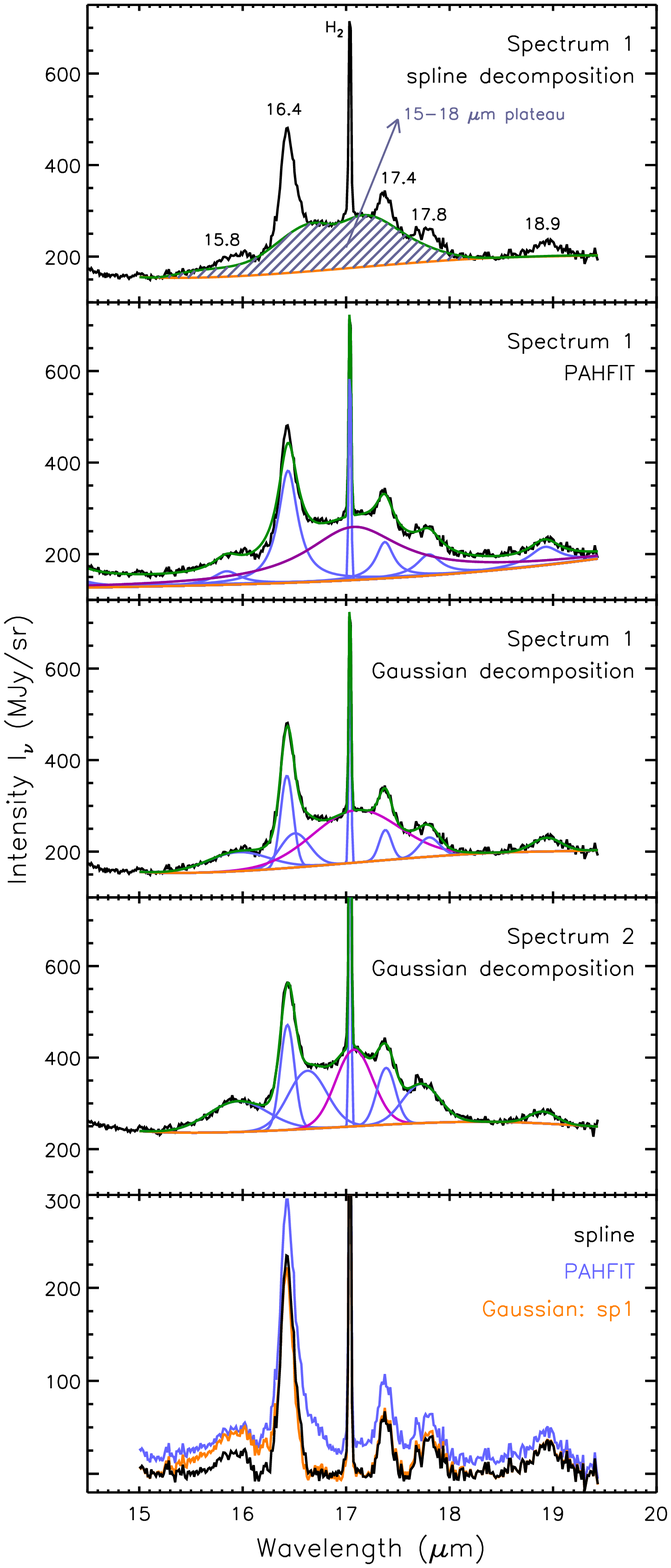}}
\caption{Typical 15-20 \mum\, spectra towards NGC\,2023. The top panel exemplifies the spline decomposition applied in this paper: the local continuum (green), global continuum (red) and the 15-18 \mum\, plateau (lined region). For comparison, the decomposition obtained with PAHFIT is shown in the second panel and the decomposition obtained by applying the Gaussian decomposition used by \citet{Sellgren:07} for their analysis of NGC7023 is shown in the middle panel. This Gaussian decomposition is also applied to a second spectrum towards NGC\,2023 (fourth panel). The colour coding is as follows: continuum (red), fit (green), individual components (blue), the broad 17 \mum\, component (PAHFIT, purple) and the broad 17.2 \mum\, component (Gaussian decomposition, magenta). The bottom panel shows the residual spectra obtained from the observed spectrum 1 by subtracting the local spline continuum, the 17 \mum\, PAHFIT component and both the 16.6 and 17.2 \mum\, Gaussian components respectively.  }
\label{fig_sp}
\end{figure}

A typical 15-20 \mum\, spectrum of NGC\,2023 reveals a rising dust continuum, a sharp H$_2$ emission line (17 \mum), an emission band at 19 \mum\, attributed to C$_{60}$ \citep{Cami:10, Sellgren:10}, a feature at 17.4 \mum\, which is a blend of a PAH band with another C$_{60}$ band and a plethora of PAH emission bands at 15.8, 16.4, 17.4 and, 17.8 \mum\,  on top of a broad underlying emission component (Fig.\,\ref{fig_sp}). 

NGC2023 contains a cluster of young stars \citep{Sellgren:83}. The YSO source D is located just outside the FOV of the south map (see Fig.\,\ref{fov}). Nevertheless, the spectra of regions close to source D show PAH emission features as well as characteristics typically for YSOs, i.e. a strong dust continuum (see the continuum map at 19.29 \mum\, in Fig.\,\ref{fig_maps_s}) and a (strong) CO$_2$ ice feature near 15 \mum. Furthermore, the background contribution to the 11.2 PAH flux is significant for a large fraction of these (ice-)spectra. We therefore excluded these spectra in the PAH and the 14.9 \mum\, continuum analysis. The 15-20 \mum\, spectrum of the YSO source C has the same spectral characteristics as the spectra across NGC\,2023 but has an enhanced surface brightness (see pixels $\sim$ (21, 6) in the continuum maps at 14.9 and 19.29 \mum\, in Fig.\,\ref{fig_maps_s}). However, we did not exclude these spectra in the analysis. 

\subsection{Continuum and fluxes}
\label{cont}
We determined the fluxes of the PAH features in the 15-20 \mum\, region by subtracting a {\it local} spline continuum shown as a green line in the top frame in Fig.\, \ref{fig_sp}. This continuum is determined by using anchor points at 14.91, 15.20, 15.50, 16.14, 16.69, 16.89, 17.16/17.19 (for the south and north maps, respectively), 17.58, 18.14/18.17 (for the north and south maps, respectively), 18.50, 19.29 and 19.36 \mum. The fluxes are then estimated by fitting a Gaussian profile to the individual emission bands. A second {\it global} continuum (red line top frame) is determined by using anchor points at 14.91, 18.50 and 19.36 \mum. In this way, a plateau underneath the individual bands is defined by the difference of both continua and is further referred to as the 15-18 \mum\, plateau. This plateau is distinct from the 15-20 \mum\, plateau observed towards \HII\, regions \citep{VanKerckhoven:plat:00} which exhibits emission over the entire 15-20 \mum\, range. This decomposition method is further referred to as the spline decomposition.

The applied decomposition of the 15-20 \mum\, region is not unique. For example, \citet{Sellgren:07} fit this region by a combination of a continuum fit (similar to our global continuum) and eight Gaussians (further referred to as the Gaussian decomposition). Similarly, a combination of Drude profiles is regularly used \citep[PAHFIT,][]{SmithJD:07}. For comparison, we applied both methods to our spectra (Fig.\, \ref{fig_sp}).
To provide sufficiently wavelength coverage for PAHFIT, we obtained the SL spectrum for the same extraction aperture as the SH spectrum. However, we did not correct for the different PSF at these wavelengths.  The PAHFIT decomposition in the 15-20 \mum\, region results in components representing the dust continuum emission \citep[which is a combination of modified blackbodies, see][]{SmithJD:07}, the H$_2$ emission and the PAH dust features including a broad component centered at 17.0 \mum. We also applied
a similar decomposition as \citet[][their fig. 8 and table 1]{Sellgren:07} to our spectra by using their Gaussian parameters as a starting value. This method results in two possible decompositions of the 15-20 \mum\, region; one dominated by a broad component peaking around $\sim$ 17.2 \mum\, (Fig.\, \ref{fig_sp}, third panel) or another one in which this 17.2 \mum\, component is much narrower resulting in broader and stronger components near 16.6 and 17.4 \mum\, (Fig.\, \ref{fig_sp},  fourth panel). Hence, in order to study the PAH bands in a consistent way, one needs to restrict the decomposition to a broad {\it or} narrow 17.2 \mum\, component. But choosing one over the other will clearly influence the derived fluxes of the Gaussian components. 

The Gaussian decomposition with the broad 17.2 \mum\, component results in very similar fluxes for the features as those obtained with the spline decomposition discussed above, except for the 15.8 and 16.4 \mum\, bands (Fig.\, \ref{fig_sp}, bottom panel). In contrast, different PAH band intensities are obtained with the Gaussian decomposition with the narrow 17.2 \mum\, component and the PAHFIT decomposition. In particular, the broad component centered around 17 \mum\, depends significantly on the decomposition method. Here we will adopt the spline decomposition (Fig.\, \ref{fig_sp}, top panel). 

\section{Data analysis}
\label{analysis}

Here we investigate i) the relationship between individual PAH emission bands and ii) the relationship between the different emission components - PAHs, H$_2$, C$_{60}$, and dust continuum - present in the 15-20 \mum\, region based upon the spatial distribution and the intensity ratios of the various emission components. We further investigate the PAH band profiles in the 15-20 \mum\, region. The implications of these results are discussed in the next section.

\begin{sidewaysfigure*}
    \centering
\resizebox{\hsize}{!}{%
\includegraphics{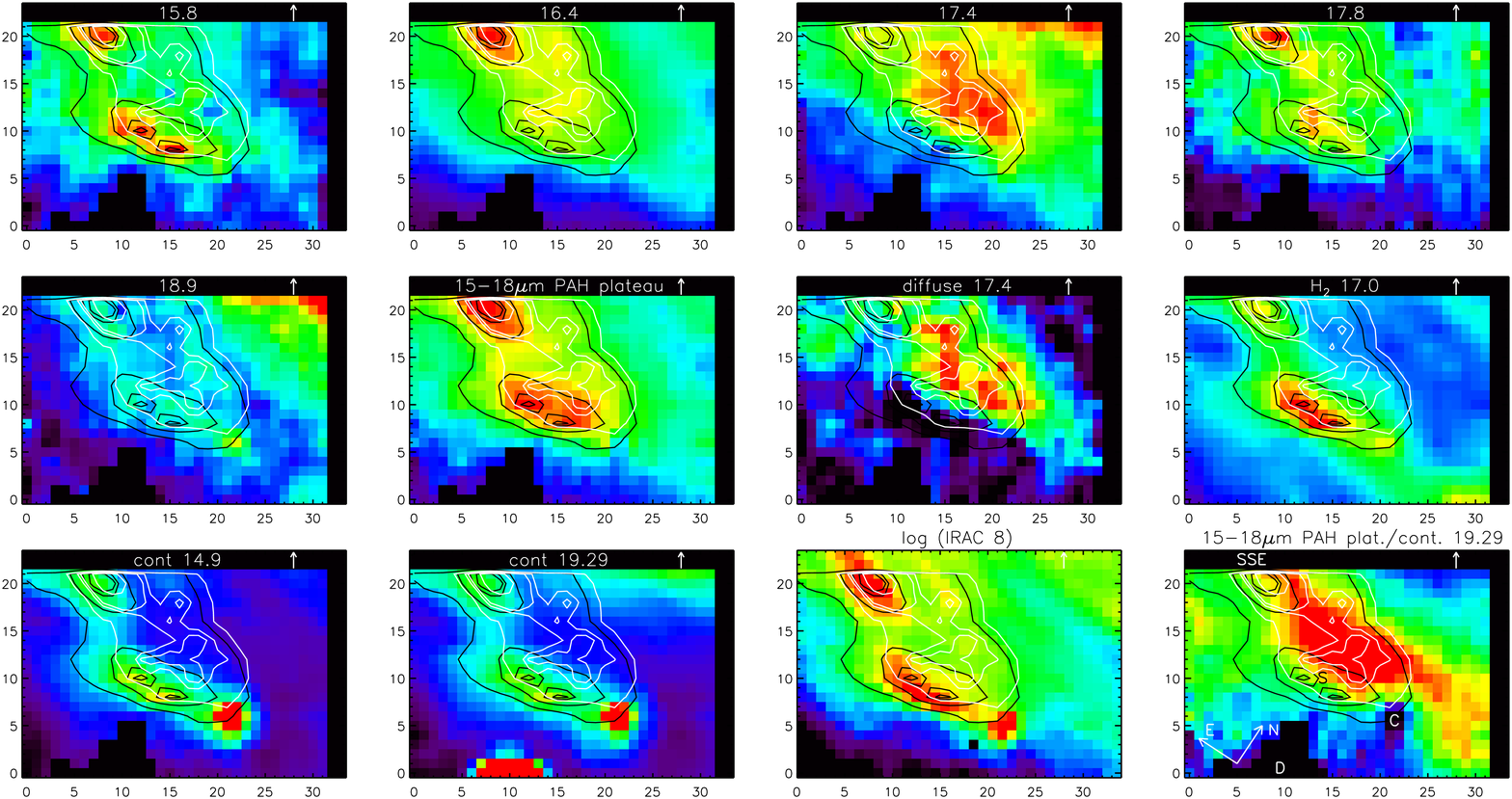}}
\caption{Spatial distribution of the emission features in the 15-20 \mum\, region in NGC\,2023 for the south map. As a reference, the intensity profiles of the 11.2 and 12.7 \mum\, emission features (Peeters et al. 2012, in prep.) are shown as contours in black and white, respectively. The white arrow in the top right corners indicates the direction towards the central star and the E-N orientation is given in the bottom right panel (see Fig.\,\ref{fov}). S refers to the south ridge, SSE to the south-southeastern ridge, C to source C and D to source D. The axis labels refer to pixel numbers. The PAH fluxes and the 14.9 \mum\, continuum flux in regions near source D excluded from the analysis are set to zero (near pixel (10,0); see Section\,\ref{sp}).}  
\label{fig_maps_s}
\end{sidewaysfigure*}

\begin{sidewaysfigure*}
    \centering
\resizebox{\hsize}{!}{%
\includegraphics{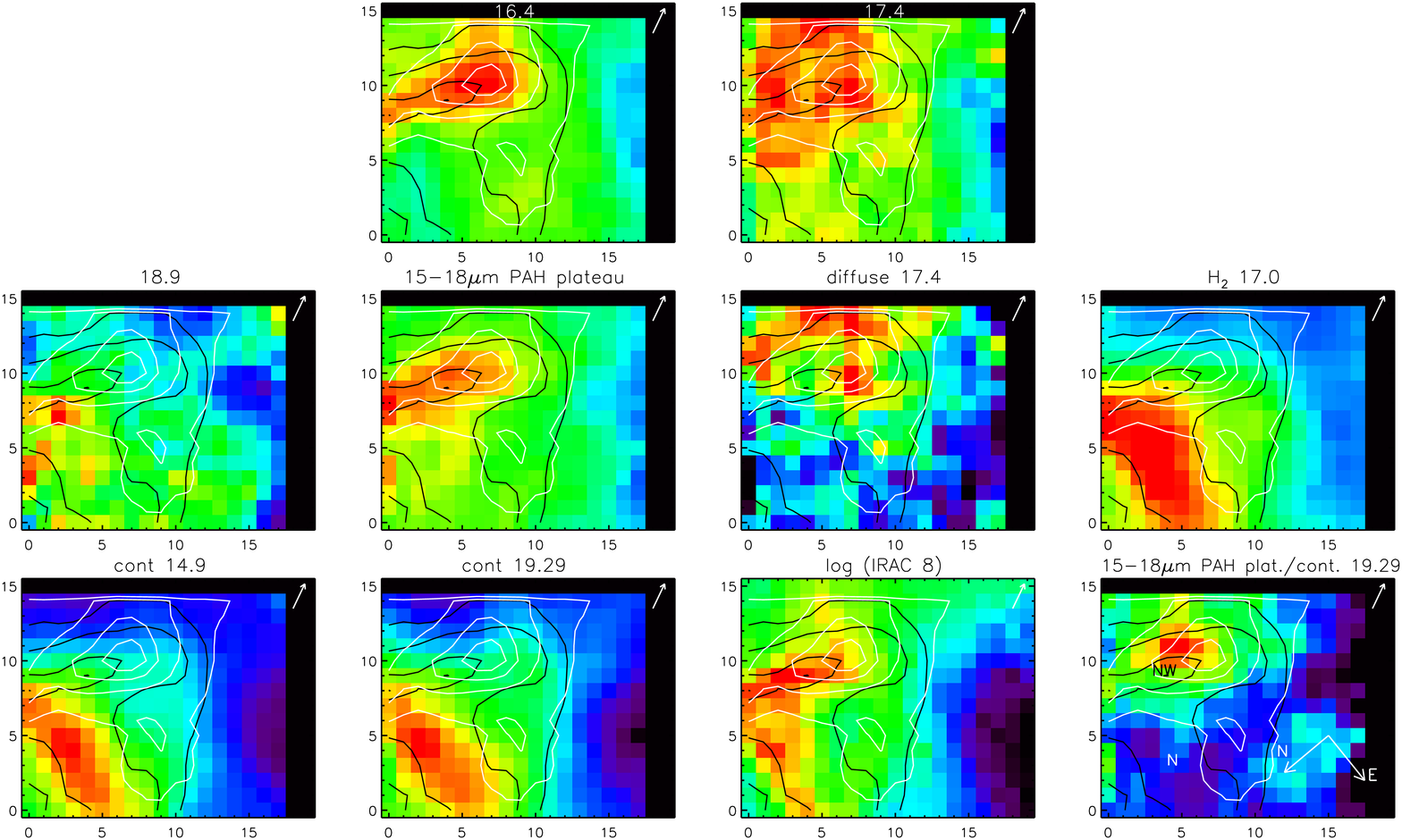}}
\caption{Spatial distribution of the emission features in the 15-20 \mum\, region in NGC\,2023 for the north map. As a reference, the intensity profiles of the 11.2 and 12.7 \mum\, emission features (Peeters et al. 2012, in prep.) are shown as contours in black and white, respectively. The white arrow in the top right corners indicates the direction towards the central star and the E-N orientation is given in the bottom right panel (see Fig.\,\ref{fov}). N refers to the north ridge and NW to the northwestern ridge. The axis labels refer to pixel numbers. }
\label{fig_maps_n}
\end{sidewaysfigure*}

\begin{figure*}[!t]
    \centering
\resizebox{14cm}{!}{%
  \includegraphics{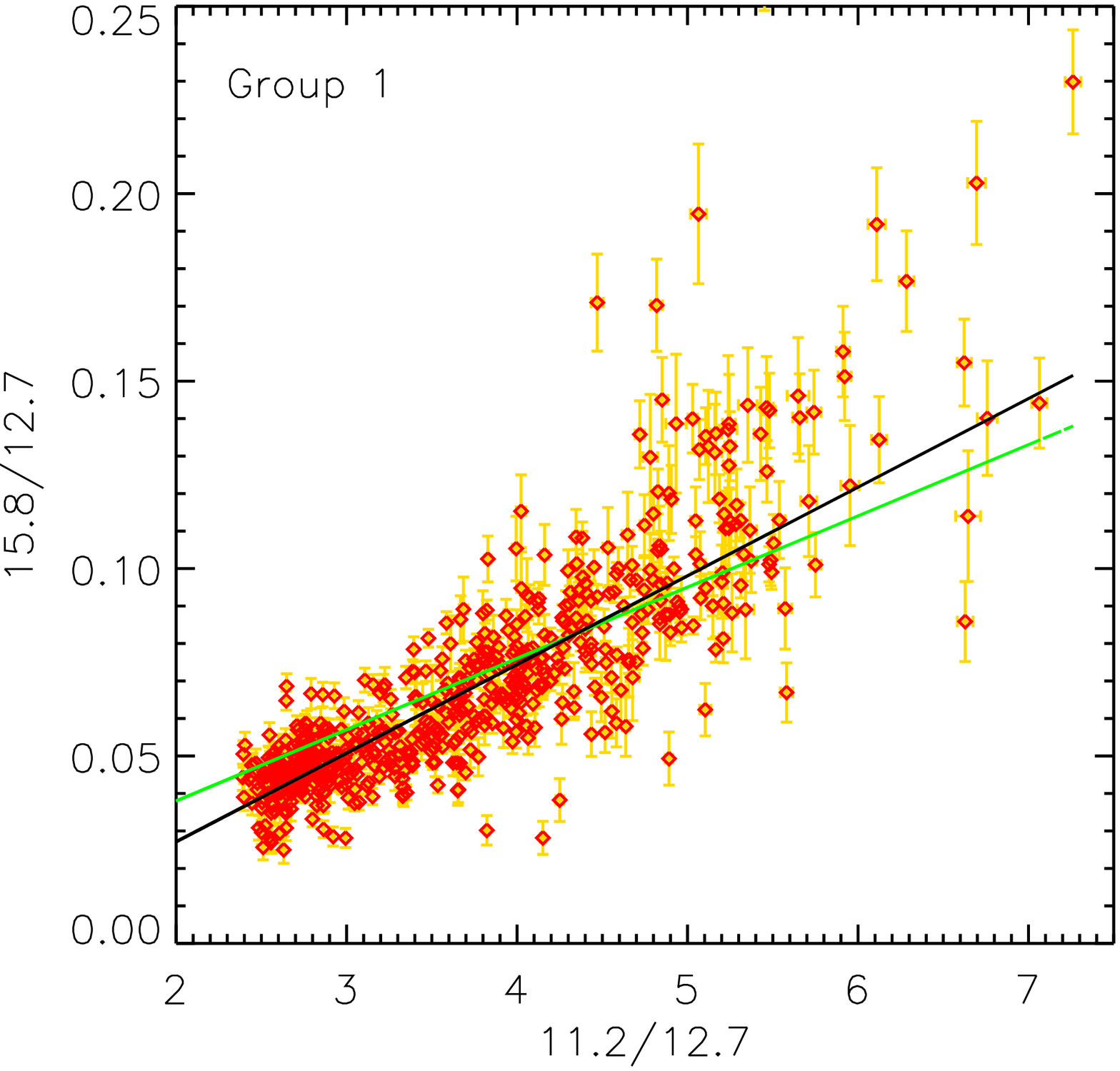}
  \includegraphics{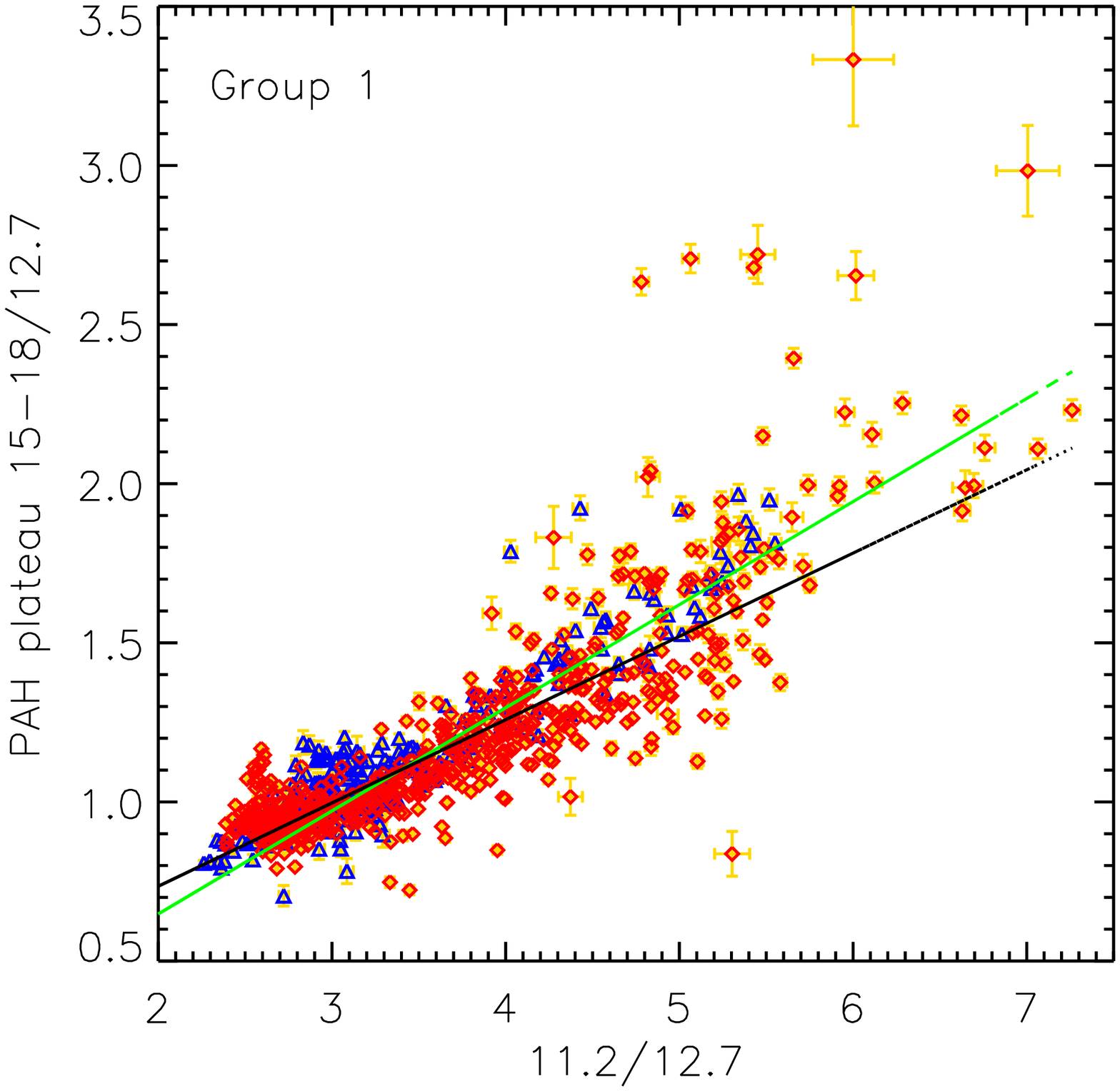}}
    \resizebox{14cm}{!}{%
   \includegraphics{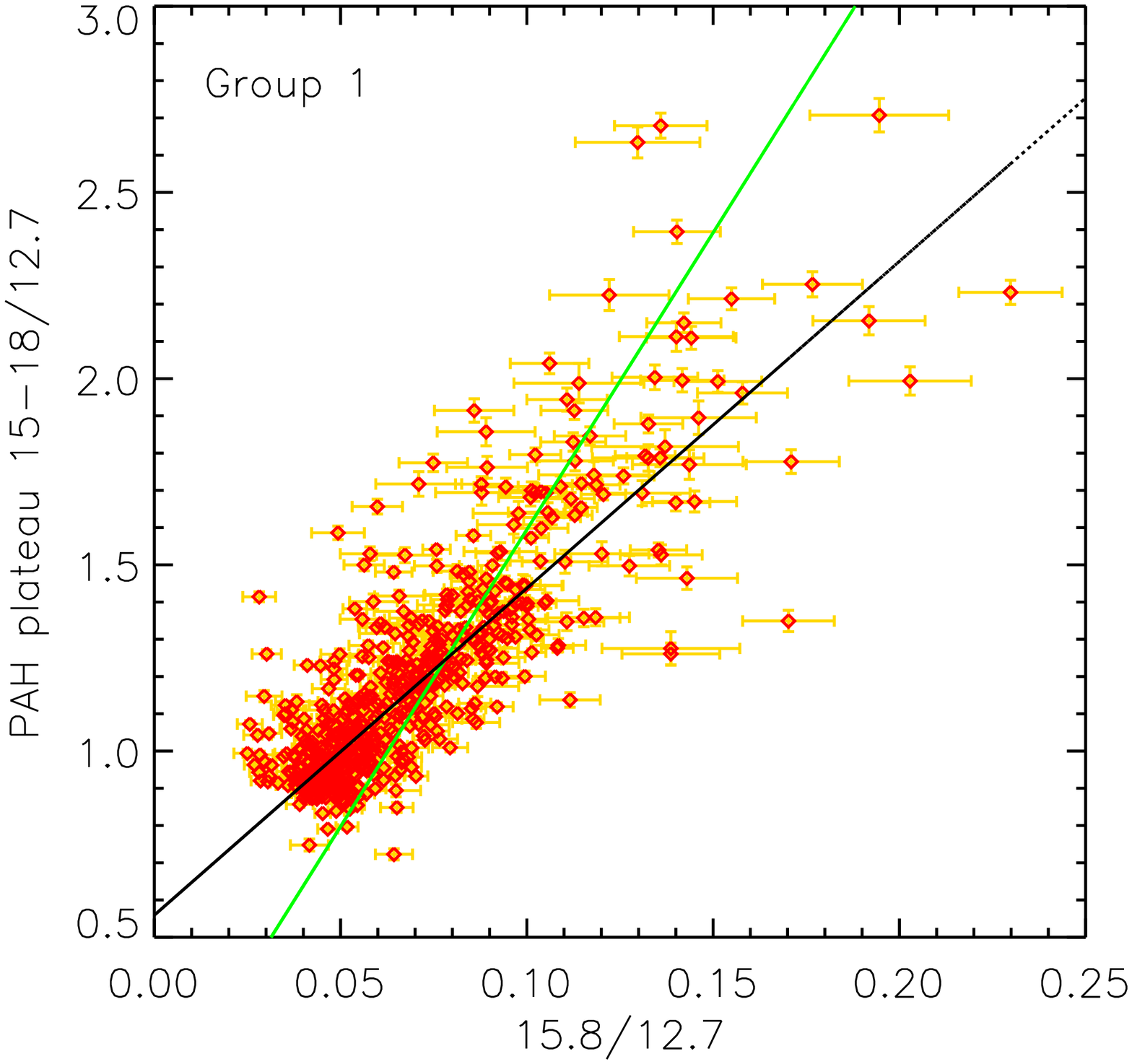}
     \includegraphics{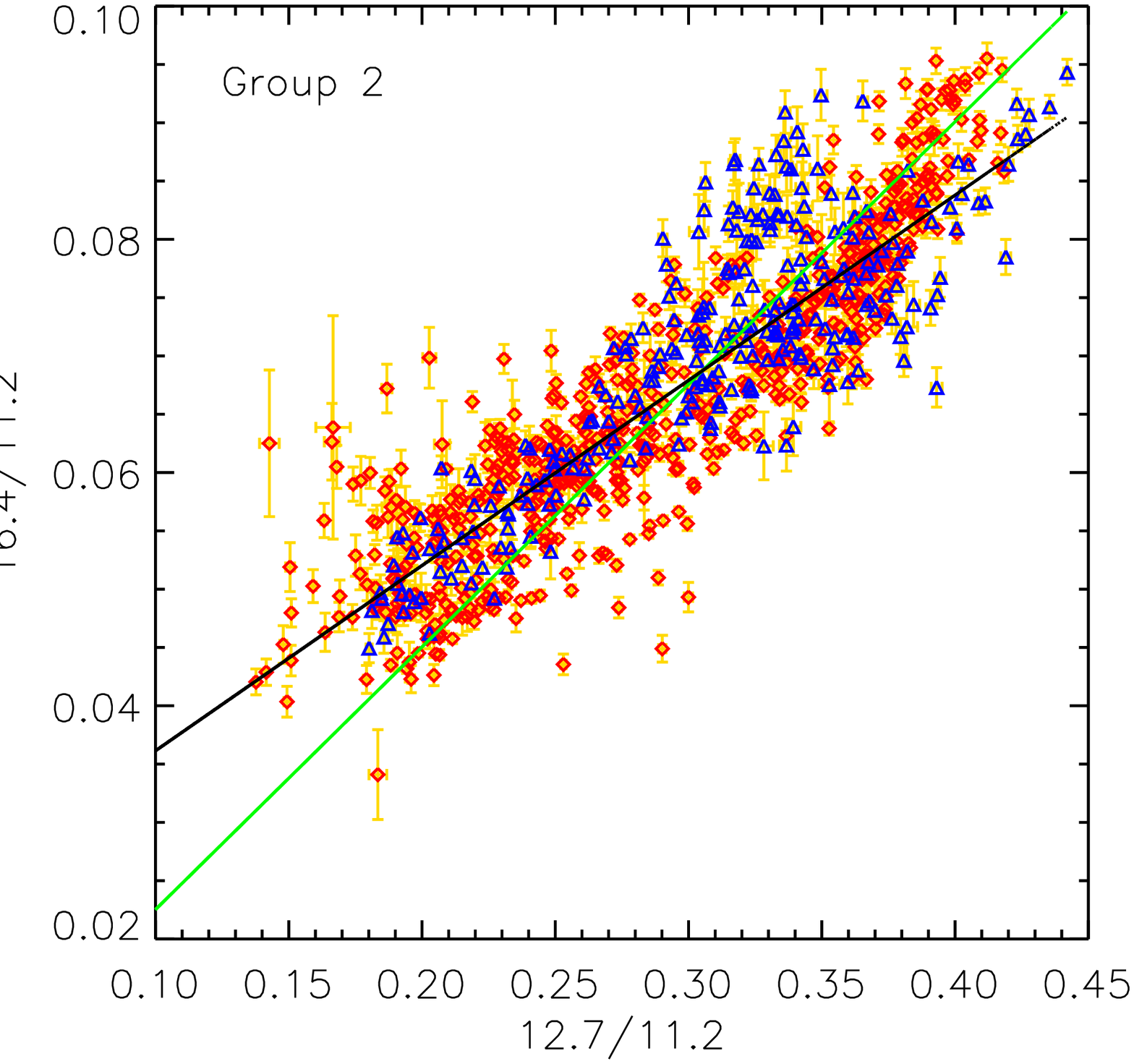}}
\caption{Correlations in PAH intensity ratios across NGC\,2023. The red squares are for the south map and the blue triangles for the north map. The fit to the data is shown as a solid black line and its parameters can be found in Table \ref{fit_parameters}. The green line represents a fit to the data forced through (0,0). }
\label{fig_correlations}
\end{figure*}

\begin{figure}[h!]
    \centering
\resizebox{7cm}{!}{%
  \includegraphics{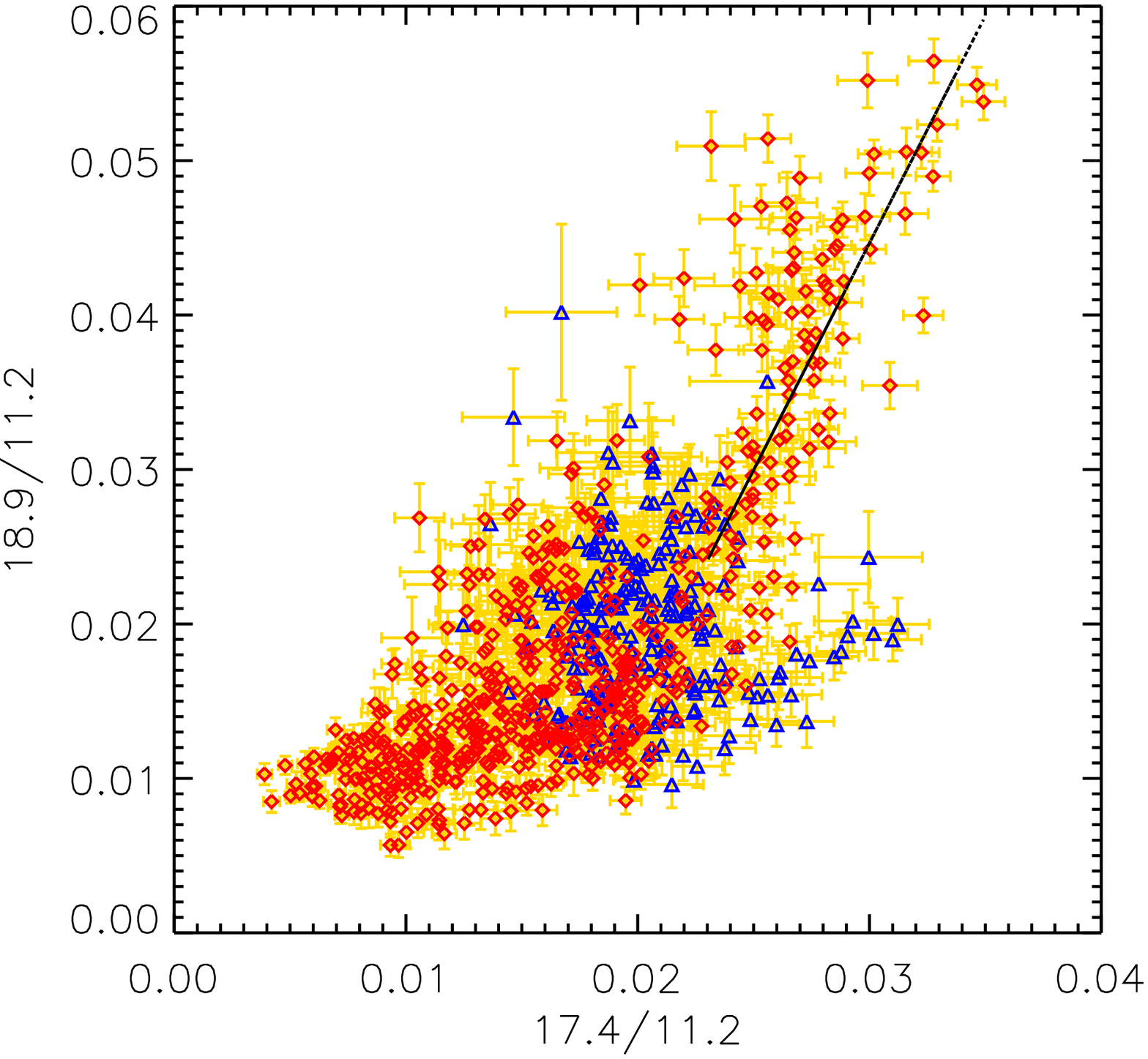}}
\caption{The 17.4 versus the 18.9 \mum\, band intensities across NGC\,2023. The red squares represent the south map and the blue triangles the north map. The fit to the data for high band intensity ratios is shown as a black line.}
\label{fig_correlations_c60}
\end{figure}

\subsection{Relationship between emission components}
\label{groups}

{\it Maps}: Figs. \ref{fig_maps_s} and \ref{fig_maps_n} show the spatial distribution of the various emission components in the 15-20 \mum\, region. In these figures, in addition to the color-coded maps for the different features, we show contours for the 11.2 and 12.7 \mum\, PAH bands for reference. These were derived from the SH data and are discussed in Peeters et al. (2012, in prep.). The north map is characterized by lower flux levels and therefore larger scatter is present in the map tracing the 15.8 and 17.8 \mum\, features and, to a lesser extent, the 18.9 \mum\, band.  Hence, when discussing the spatial distribution of the 15.8 and 17.8 \mum\, bands, we restrict ourselves to the south map. 

{\it Feature correlations}: To investigate possible variations within the PAH spectrum itself, we need to exclude the influence of PAH abundance and column density. This is achieved by normalizing the fluxes to another band in this region. Here we start with the 11.2 \mum\, PAH band, which is attributed to solo CH out-of-plane bending vibration of large, neutral PAHs \citep{Hony:oops:01, Bauschlicher:vlpahs1, Bauschlicher:vlpahs2}. To look for possible correlations with the 11.2 \mum\, band, we also apply a normalization to the 12.7 \mum\, PAH band which is attributed to duo and trio CH out-of-plane bending vibration of large PAHs \citep{Hony:oops:01, Bauschlicher:vlpahs1, Bauschlicher:vlpahs2}.  Both the 11.2 and 12.7 \mum\, PAH fluxes are obtained from the SH data using a spline continuum \citep[see e.g. Fig. 1 of][]{Hony:oops:01} and are discussed in Peeters et al. (2012, in prep.).  A correlation is often found between the normalized intensity ratios of PAH bands in this region as these bands arise from a family of PAH molecules which all have the same basic structural unit. However, variations in charge states, excitation level, edge structures, etc., exist within the PAH family which can tighten as well as loosen correlations between the normalized intensity ratios of specific PAH bands. For example, the well known tight correlation between the 6.2 and 7.7 \mum\, PAH bands is attributed to the common charge state of these bands \citep[e.g.][]{Hony:oops:01, Galliano:08}. Figs. \ref{fig_correlations} and \ref{fig_correlations_c60} show some of the intensity correlations found in NGC\,2023. The correlation coefficients and the fit parameters are given in Table \ref{fit_parameters}. While maps smooth out the detailed, fine grained variations, allowing one to see the big picture, these correlations highlight subtle differences.

{\it Results:} 
The following trends are derived from the south spectral maps shown in  Fig.~\ref{fig_maps_s}.  The 11.2 \mum\, feature, the 15.8 \mum\, feature and the 15-18 \mum\, plateau show very similar spatial morphology with distinct peaks at the S and SSE positions in the IRAC map. In contrast, the distribution of the 12.7, 16.4, 17.4, and 17.8 \mum\, features are displaced towards the star.  The 12.7 and 16.4 \mum\, bands have very similar spatial behavior; both peak at the SSE position and show a broad, diffuse plateau NW of the line connecting the S and SSE ridges.  The 17.4 \mum\, feature also presents a strong, diffuse plateau NW of that line, but does {\it not} show a peak at the SSE position as does the 12.7  and 16.4 \mum\, bands. In addition, the 17.4 \mum\, feature has a separate peak in the upper right of the field, coinciding with the peak in the 18.9 \mum\, band.  The 17.8 \mum\, band peaks primarily on the SSE position, overlapping with the peak of both the 11.2 and 12.7 \mum\, bands but also shows some enhanced emission slightly offset from the 11.2 \mum\, peak near the S ridge, though not as much as the 12.7 and 16.4 \mum\, bands.  The 18.9 \mum\, band is clearly in a class by itself, peaking closer to the illuminating star than any of the other features considered here. Interestingly, a second peak is evident in the 18.9 \mum\, spatial distribution at source C. Note also that the 18.9 \mum\, band does show (weak) emission tracing the S and SSE ridges. The spatial behavior of the 14.9 and 19.29 \mum\, continuum intensities align closely with the H$_2$ emission distribution, all lying along the SE portion of the 11.2 \mum\, contours, while clearly avoiding regions with strong 12.7 \mum\,  emission. The H$_2$ map shows, however, additional emission towards the west.
 
The similarity in distribution of the 11.2  \mum\, and 15-18 \mum\, plateau is also evident in the north map (Fig. \ref{fig_maps_n}): both maps peak on the NW ridge and extend to include the N ridge. There is too much scatter in the weak 15.8 \mum\, PAH band to warrant comparison. In addition, the 16.4 \mum\, PAH band has a spatial distribution very similar to that of the 12.7 \mum\, band. However, unlike the situation at the SSE ridge where the 11.2, 12.7 and 16.4 \mum\ bands all peak in the same place, the 12.7 and 16.4 \mum\, bands are clearly displaced towards the south with respect to the 11.2 \mum\, emission in both the peak emission and the N ridge extension. The 17.4 and 18.9 \mum\, bands exhibit more scatter in the north map. Nevertheless, the 17.4 \mum\, band seems to peak towards the NW ridge and further west of this ridge. In contrast, the 18.9 \mum\, band seems to trace the N ridge and not the NW ridge while also showing secondary emission in the south. The spatial behavior of the 14.9 and 19.29 \mum\, continuum emission and the H$_2$ emissions are quite distinct from the PAH emission in the north map as they do not track the 11.2 \mum\, contours as closely and only trace the N ridge and not the NW ridge. 

This morphology suggests that the features can be viewed as falling into the following groups: 1) the 11.2, 15.8 \mum\, bands and the 15-18 \mum\, plateau; 2) the 12.7, and 16.4 \mum\, bands; 3) the 17.4 \mum\, diffuse component;  4) the 17.8 \mum\, band; 5) the 18.9 and (some of) the 17.4 \mum\, features; 6) the underlying continuum producing the emission measured here at 14.9 and 19.2  \mum; and 7) the H$_2$ emission.

The correlation plots shown in Fig. \ref{fig_correlations} support grouping the 11.2, 15.8 \mum\, bands and the 15-18 \mum\, plateau into a single group. Indeed, a very tight correlation is found in our data set between the 11.2/12.7 PAH intensity ratio and the 15-18 \mum\, plateau/12.7 PAH intensity ratio.  Weaker trends, although still very tight, are seen between the 11.2/12.7 and 15.8/12.7 PAH intensity ratios and the 15.8/12.7 and 15-18 \mum\, plateau/12.7 PAH intensity ratios. In addition, the normalized 16.4 \mum\, PAH intensity clearly correlates with that of the 12.7 \mum\, PAH band; their correlation coefficient is the highest found in our data, i.e. 0.889.

\begin{table}
\small
\caption{\label{fit_parameters} Parameters of the fit to the observed correlations and their Pearson's correlation coefficients.}
\begin{center}
\begin{tabular}{c|rrr}
 PAH intensity ratios & correlation & A$^1$ & B$^1$ \\
y vs. x & coefficient  & & \\
       \hline 
       \hline
 & &  \\[-5pt]
15.8/12.7 vs. 11.2/12.7  & 0.834 & -0.02  & 0.02 \\
plat./12.7$^2$ vs. 11.2/12.7  & 0.867 & 0.21    & 0.26\\ 
plat./12.7$^2$ vs. 15.8/12.7  & 0.837 & 0.56    & 8.79 \\
16.4/11.2 vs. 12.7/11.2 & 0.889 & 0.02 & 0.16 \\
\end{tabular} 
\end{center}
$^1$ Parameters of the fit: y = A + Bx\\
$^2$ Plat. refers to the 15-18 \mum\, plateau.\\
\end{table}

As shown in Fig. \ref{fig_correlations_c60}, the 17.4 and 18.9 \mum\, bands do not exhibit an overriding linear correlation with each other as do the bands in Fig. \ref{fig_correlations}. Instead, the 18.9 \mum\, band does not correlate with the 17.4 \mum\, band for low 18.9/11.2 intensity ratios ($<$ 0.02) and shows a linear correlation with the 17.4 \mum\, band for strong band ratios. This is consistent with the assignment of the 18.9 \mum\, band to C$_{60}$ and the 17.4 \mum\, feature to being a blend of a PAH emission band with a C$_{60}$ band \citep{Cami:10, Sellgren:10}. Therefore,  the 17.4 \mum\, band was decomposed into these two components to estimate the spatial distribution of the PAH contribution to this band.  This is referred to as the diffuse 17.4 \mum\, band. We calculated the average ratio of the 18.9/17.4 band intensity in the upper right corner of the {\it south} map (where the 18.9 \mum\, band peaks) and used this value to subtract the C$_{60}$ contribution to the 17.4 \mum\, band to obtain the "diffuse 17.4 \mum\, band" for both the south and north map (see Fig. \ref{fig_maps_s} and \ref{fig_maps_n}). Note that this method assumes a constant ratio between the 17.4 and 18.9 \mum\, C$_{60}$ emission (of 0.6, similar to what is observed e.g. in Tc1 \citep{Cami:10}) and only provides a first order approximation to the diffuse 17.4 \mum\, band distribution. Nevertheless, this lends support to our earlier suggestion that the PAH component in the 17.4 \mum\, feature does not spatially coincide with the 11.2 or 12.7 \mum\, PAH emission.

\begin{figure}[t!]
    \centering
\resizebox{8cm}{!}{%
  \includegraphics{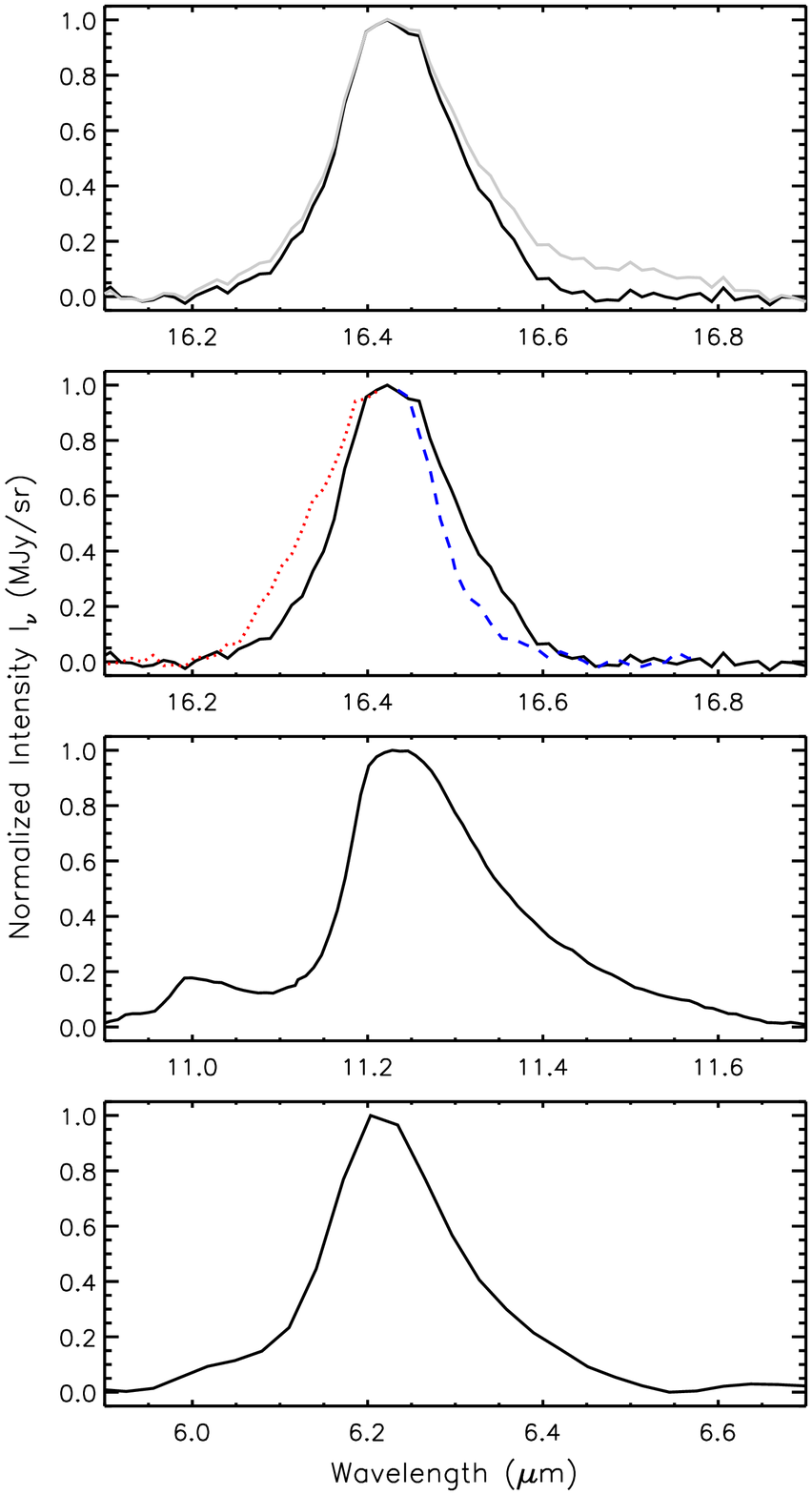}}
\caption{The asymmetric 16.4 \mum\, band (top panel).  Two profiles are shown to assess the  influence of the continuum determination on band shape: a spline fit to the continuum points discussed in Sect. \ref{cont} (black solid line) and a spline fit to the same continuum points excluding that at 16.69 \mum\, (grey solid line). The asymmetry of the black profile is assessed in the second top panel: the dotted red line shows the red wing mirrored along a vertical position at the peak position and the dashed blue line represents the mirrored blue wing.  For reference, the asymmetric 6.2 and 11.2 \mum\, band profiles are shown in the bottom and middle panel respectively.  Note that the 6.2 band profile is obtained from the IRS-SL observations and the 11.2 band profile from the IRS-SH observations.} 
\label{fig_prof}
\end{figure}

\begin{figure}[t!]
    \centering
\resizebox{8cm}{!}{%
  \includegraphics{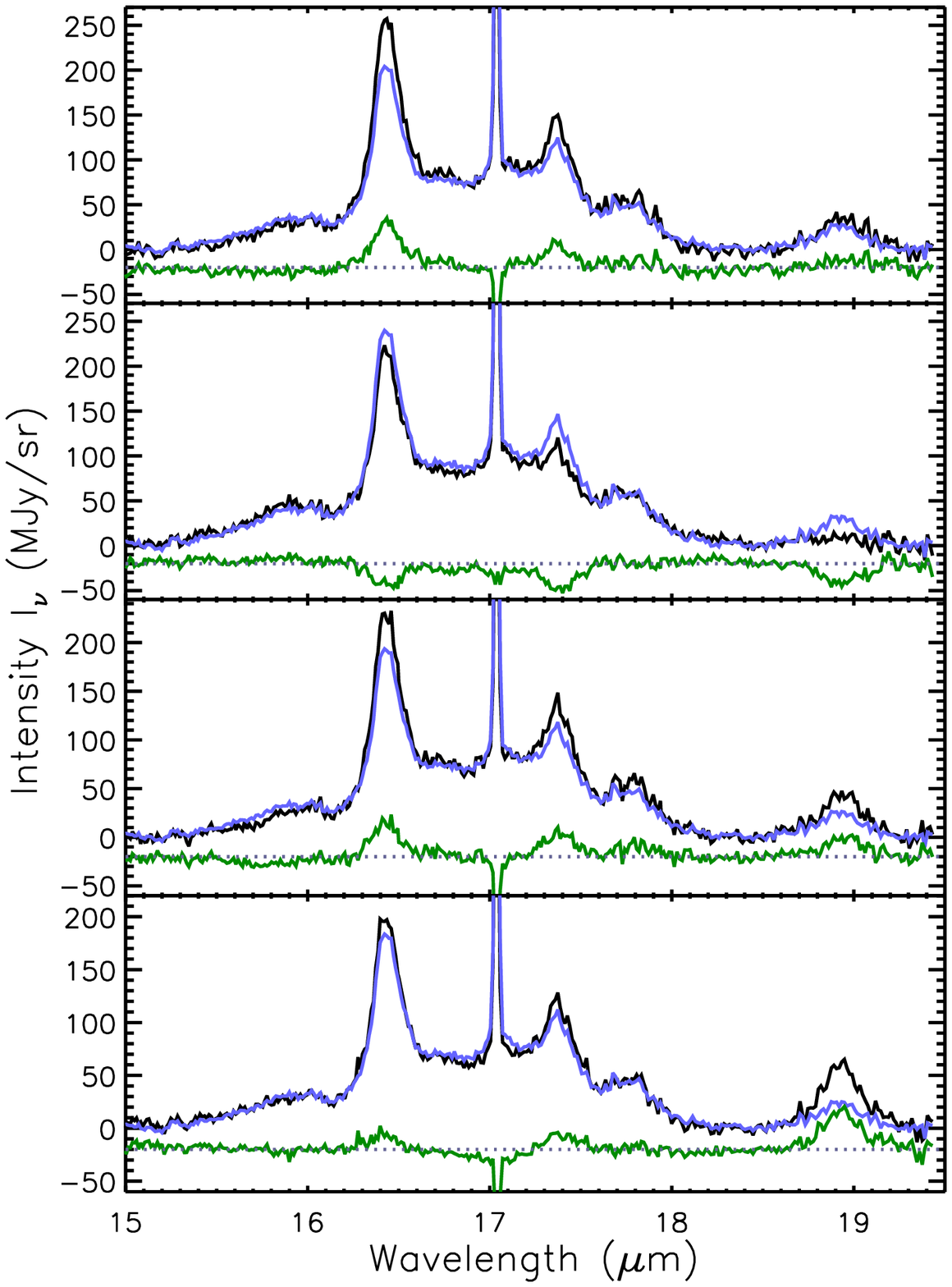}}
\caption{The 15-18 \mum\, plateau. Shown are spectra from different pixels (black), the average 15-18 \mum\, plateau scaled such that its flux matches the 15-18 \mum\, plateau flux of the black spectra (blue) and, the residuals (green). The second panel shows the extend of the variation seen in the 15-18 \mum\, plateau profile. Note that the strength of the features varies independently of the 15-18 \mum\, plateau and of each other.}
\label{fig_plat}
\end{figure}

\subsection{Band profiles}

We investigate the band profiles in the spectrum from the total FOV of the south map to minimize the noise and apply the same continuum determination as discussed in Sect. \ref{cont}. Within the uncertainty, the 15.8, 17.4, 17.8, and 18.9 \mum\, bands are well fitted with a Gaussian. Hence, no shift is found in the central wavelength of the 17.4 $\mu$m feature as the 17.4/18.9 intensity ratio varies. In contrast, the 16.4 \mum\, PAH feature is slightly asymmetric to the red (Fig. \ref{fig_prof}). To assess the influence of the continuum on the degree of asymmetry, three different continua were constructed and the resulting 16.4 \mum\, bands compared. First, the continuum point near 16.69 \mum\, may be considered to be part of the red wing of the 16.4 \mum\, band. Hence, it is omitted and a spline is fitted to the remainding continuum points. The other two ways involve fitting a straight line to  continuum points at 16.14 \mum\, and 16.69 or 16.89 \mum\, respectively. There is little difference between the spline fit and the straight line but the omission of the continuum point at 16.69 \mum\, clearly amplifies the red wing of the band and thus the asymmetry (Fig. \ref{fig_prof}). Comparison of the 16.4 \mum\, band profile with the well-known asymmetric 6.2 and 11.2 \mum\, PAH bands shows that its asymmetry is less pronounced (Fig. \ref{fig_prof}).  

In order to investigate possible variations in the profile of the 15-18 \mum\, plateau, we compared the 15-18 \mum\, plateau of each pixel with the average 15-18 \mum\, plateau (Fig. \ref{fig_plat}). Overall, we conclude that the profile of the 15-18 \mum\, plateau does not vary, despite a possible change in intensity of the PAH bands located on top of this plateau. Only in a handful of cases do we see slight deviation from the average. Hence, the 15-18 \mum\, plateau profile does {\it not} depend on the individual PAH bands as determined in this paper.

\section{Discussion}
\label{discussion}

\subsection{Correlation studies}
In a previous detailed study of the PAH emission features in the 15-20 \mum\ region, \citet{Boersma:10} focused on the spectra from very different astronomical environments, not the spatial behavior in an extended object as considered here.  Analyzing the spectra from some 10 objects, \citet{Boersma:10} showed that the 16.4 \mum\, band correlated with the mid-IR 6.2 and 7.6/7.8 \mum\ bands and that the total 15-20 \mum\, PAH emission correlated with the 11.2 \mum\ band.  No further connections were found between the 15.8, 16.4, 17.4, 17.8 and 18.9 \mum\, bands or between the spectral characteristics of the 15-20 \mum\, emission and the mid-IR PAH modes. These authors then conclude that these bands must be carried by independent molecular species or classes of species. The results presented in the previous section both support and amplify these conclusions. The strong correlation of the 16.4 and the 6.2 \mum\, band found by \citet{Boersma:10} is confirmed by our results as both the 16.4 and the 12.7 \mum\, PAH bands strongly correlate with each other and it is well known that the 12.7 \mum\, correlates well with the 6.2 \mum\, band \citep[][Peeters et al. 2012, in prep.]{Hony:oops:01}.  In contrast with \citet{Boersma:10}, we find a correlation between the 15.8 and 11.2 \mum\, bands. The 15.8 \mum\, band is very weak and only detected in a few sources in their sample. These sources follow the correlation found in this paper.

\subsection{The 15-18 \mum\, plateau}

As discussed in Sect. \ref{cont}, different spectral decompositions have been applied in the literature. This is in particular true for the treatment of the broad underlying plateaus. They are either considered to be part of the PAH bands themselves which are then fitted by Lorentzians or Drude profiles \citep[e.g.][]{Boulanger:lorentz:98, SmithJD:07}, or they are considered to be independent of these PAH bands \citep[e.g.][this paper]{Hony:oops:01, Peeters:prof6:02}. The spatial maps and the band correlations clearly indicate that there are significant differences in the spatial distribution between the 15-18 \mum\, plateau and most PAH bands. These spatial differences suggest a different carrier for the plateau and the individual PAH bands (with the exclusion of the 15.8 \mum\, band), a finding consistent with earlier suggestions for the plateaus underlying the prominent mid-IR bands between 5 and 13 \mum\, \citep{Bregman:orion:89, Roche:orion:89}.

\subsection{The 15-20 \mum\, emission components}

The emission features between 15-20 \mum\,  can be collected into seven groups as discussed in Section \ref{groups}. Here we discuss these in terms of the specific characteristics that set each group apart and make them particularly sensitive to the local physical conditions. 
    
{\it Group 1: The 11.2, 15.8 \mum\, bands and the 15-18 \mum\, plateau.}  Since the 11.2 \mum\ band is associated with the solo CH out-of-plane vibration of large, neutral PAHs \citep[e.g.][]{Allamandola:modelobs:99, Hony:oops:01, Galliano:08, Bauschlicher:vlpahs1, Bauschlicher:vlpahs2}, the 15.8 \mum\, band and the 15-18 \mum\, plateau must arise from species that are favored by the same conditions that favor this subset of the PAH population.
In contrast with expectations, the 15-20 \mum\, region shows little systematic dependence on PAH class or molecular structure as present in the NASA Ames PAH IR spectroscopic database \citep{Bauschlicher:10, Boersma:10, Ricca:10}, suggesting that molecular structure is not the main origin for the observed correlations. Keep in mind that, to date, the database is biased to smaller PAHs and the astronomical emission at these longer wavelengths is dominated by larger PAHs. Thus, as knowledge of larger PAH spectroscopy increases, this suggestion should be revisited. In addition, ionization only has a moderate influence on the PAH emission in the 15-20 \mum\, region \citep{VanKerckhoven:plat:00, Boersma:10}: it slightly influences the relative strength of the emission bands but not their peak position. Based on theoretical and experimental PAH spectra, one can therefore not assign a particular charge state to these individual bands, unlike for the mid-IR bands. 
Therefore, we suggest that large, neutral PAHs are also responsible for the 15.8 \mum\, band and 15-18 \mum\, plateau. PAH clusters, VSGs (Very Small Grains) or small HAC particles may also contribute to the 15-18 \mum\, plateau. However, the spread of the correlations in Fig. \ref{fig_correlations} shows there is not a one-to-one correspondence between the carriers of these different emission features.  These differences reflect minor changes in the relative populations of the different band carriers, perhaps arising from slightly different molecular structures, excitation levels, and so on.  

{\it Group 2: The 12.7 and 16.4 \mum\, bands. } Similarly, since the 16.4 and 12.7  \mum\ bands correlate with the 6.2 and 7.7 \mum\ PAH cation bands \citep[][Peeters et al. 2012, in prep.]{Hony:oops:01, Boersma:10},  the 12.7 and 16.4 \mum\, bands must arise from species that are co-spatial with the cationic portion of the emitting PAH population. Again, as with Group 1, the spread of the correlations in Fig. \ref{fig_correlations} shows there is not a one-to-one correspondence between these different carriers. However, in contrast with Group 1, we cannot propose molecular structure or charge as the main origin for these bands. Indeed, laboratory and theoretical PAH spectra indicate that the 12.7 and 16.4 \mum\, bands are sensitive to the molecular edge structure: the former being due to duo and trio CH groups \citep{Hony:oops:01, Bauschlicher:vlpahs1, Bauschlicher:vlpahs2} and the latter being systematically present in PAHs with pendent rings \citep{Moutou:lissabon:98, VanKerckhoven:plat:00, Peeters:plat:04, Boersma:10} and large PAHs with pointy edges \citep{Ricca:10}. Laboratory and theoretical PAH spectra also indicate that both neutral and ionized PAHs can contribute to the 12.7 and 16.4 \mum\, emission \citep[][]{VanKerckhoven:plat:00, Hony:oops:01, Bauschlicher:vlpahs1, Bauschlicher:vlpahs2, Boersma:10} and hence do not allow, by themselves, the assignment of these bands to a single charge state. Similarly, blind signal separation of the PAH emission across PDRs cannot assign a single charge state to the 12.7 \mum\, band (the 16.4 \mum\, band is not analysed by this method). Indeed, this method revealed three spatially different components, referred to as PAH$^0$, PAH$^+$ and VSG, which are attributed to respectively neutral PAHs, ionized PAHs, and PAH clusters \citep[e.g.][]{Rapacioli:05, Berne:07, Rosenberg:11}. The 12.7 \mum\, band is present in both the PAH$^0$ and PAH$^+$ components. However, the relative contribution of the 12.7 \mum\, band in the PAH$^0$ and PAH$^+$ components seems to vary for different decomposition methods and sources \citep{Rapacioli:05, Berne:07, Joblin:08}. Hence, further investigation is warranted. 

{\it Group 3: The 17.4 \mum\, diffuse component.}  This paper confirms the results by \citet{Sellgren:10} that the 17.4 \mum\, emission band is composed of emission due to C$_{60}$ and PAHs (i.e., the diffuse 17.4 \mum\, component). In addition, \citet{Sellgren:10} report a similar spatial distribution between the diffuse 17.4 \mum\, emission and the 16.4 \mum\, PAH band in the reflection nebulae NGC7023. In contrast, in NGC\,2023, the spatial distribution of the diffuse 17.4 \mum\, emission is quite distinct from that of the 16.4 \mum\, band.  In particular, in the north map, the diffuse 17.4 \mum\, emission is strong close to the peak 16.4 \mum\, emission as well as west of the NW ridge. In the south map, it is located closer towards the illuminating star compared to the 11.2, 12.7 and 16.4 \mum\, PAH emission and does not peak in the SSE ridge where the 11.2, 12.7 and 16.4  \mum\, PAH bands peak.  Perusal of the NASA Ames PAH IR spectroscopic database \citep{Bauschlicher:10} reveals that large compact PAHs systematically show emission near 17.4 \mum\, and that PAH charge only has moderate influence on the relative intensities and no influence on the peak position of the bands in the 15-20 \mum\, region \citep{Boersma:10}.  We therefore conclude that the diffuse 17.4 \mum\, band is due to doubly ionized PAHs and/or a subset of PAH cations that favor more harsh conditions compared to those responsible for the 6.2 \mum\, PAH bands, such as for example dehydrogenated cations. This possible assignment will further be explored by comparison to the PAH emission at shorter wavelengths in a forthcoming publication (Peeters et al. 2012, in prep.).

{\it Group 4: The 17.8 \mum\, band.} The 17.8 \mum\, emission exhibits spatial characteristics between those of the 11.2 and 12.7 \mum\, bands. With the former spatial distribution being attributed to neutral PAHs and the latter being co-spatial with that of the PAH cations, the 17.8 \mum\, band likely originates from both neutrals and cations (with both having similar intrinsic strengths for this band). 

{\it Group 5: The 18.9 and (some of) the 17.4 \mum\, feature.}  These two bands have recently been assigned to C$_{60}$ \citep{Cami:10, Sellgren:10}.  The C$_{60}$ bands show a distinct spatial distribution. In the south, they reach their peak emission intensities in regions closer to the exciting star than are the PAH bands, similar to what is observed towards NGC7023 \citep{Sellgren:10}. In addition, secondary peaks in emission are found near source C, a YSO, and where the PAH emission peaks. In contrast, in the north, the peak in the C$_{60}$ distribution seems to be farther away from the illuminating source (no other UV sources are present near this position) and linked to the H/H$_2$ transition zone.  This different behavior in the C$_{60}$ emission may well provide important clues to the formation and evolution of C$_{60}$ in the ISM. This will be further explored when the full PAH spectra will be analyzed (Peeters et al. 2012, in prep.). 

{\it Group 6: The continuum producing the emission at 14.9 and 19.2  \mum.} This emission component peaks further into the filament with respect to the PAH emission. This is consistent with the results of the blind signal separation discussed above: the VSG component, attributed to PAH clusters, carries the mid-IR continuum \citep[e.g.][]{Rapacioli:05, Berne:07}. As it is located deeper in the PDR compared to the PAH emission, these authors suggested the destruction of PAH clusters by UV photons and subsequent formation of PAHs. As mentioned above, PAH clusters may also contribute to the 15-18 \mum\, plateau, which is slightly offset compared to the dust continuum emission. This suggest that the carriers of these emission components slightly differ.

{\it Group 7: The H$_2$ emission.} For the analysis of the H$_2$ emission present in this data set, we refer to \citet{Fleming:10} and \citet{Sheffer:11}.

\section{Conclusions}
\label{conclusion}

We studied the emission in the 15-20 \mum\, region towards the reflection nebula NGC\,2023 by analysing two spectral maps obtained with Spitzer's Infrared Spectrograph (IRS), short-wavelength high-resolution mode. We observed PAH emission bands at 15.8, 16.4, 17.4, and 17.8 \mum, a broad PAH plateau between 15-18 \mum, C$_{60}$ emission at 17.4 and 18.9 \mum, and H$_2$ emission at 17.0 \mum\, superposed on a dust continuum. \\

We found distinct spatial distributions for these emission components and found tight intensity correlations between some emission components. Based on these results, we collected the emission components in seven groups and discussed their specific characteristics. 

{\it Group 1: the 11.2, 15.8 \mum\, PAH bands and the 15-18 \mum\, PAH plateau.} 
We attributed this group to large, neutral PAHs. 

{\it Group 2: the 12.7 and 16.4 \mum\, PAH bands.} Compared to Group 1, these bands are displaced towards the illuminating star.  
We concluded that they must arise from species that are co-spatial with the cationic portion of the emitting PAH population.
  
{\it Group 3: the 17.4 \mum\, diffuse component.} We estimated the PAH contribution to the 17.4 \mum\, band and found that its spatial distribution is displaced towards the illuminating star compared to Groups 1 and 2. We assigned this PAH band to doubly ionized PAHs and/or a subset of the cationic PAH population, such as for example dehydrogenated PAHs. 

{\it Group 4: the 17.8 \mum\, PAH band.} This band has a spatial distribution in between that of Groups 1 and 2. We suggested this band arises from both neutral and ionised PAHs. 

{\it Group 5: the 18.9 and (some of) the 17.4 \mum\, features.} The C$_{60}$ emission shows a spatial distribution distinct from that of the PAH emission, consistent with \citet{Sellgren:10}.  Moreover, the spatial behavior of C$_{60}$ towards NGC2023 is different in the south and north. Specifically, the C$_{60}$ emission is located closer to the illuminating star compared to the PAH and H$_2$ emission in the south while in the north, it seems to be associated to the H/H$_2$ transition. 

{\it Group 6: the underlying continuum producing the emission measured here at 14.9 and 19.2 \mum.} The continuum emission is strongest deeper into the PDR compared to the PAH emission. This is consistent with the VSG component attributed to PAH clusters by, for example, \citet{Rapacioli:05} and \citet{Berne:07}.

{\it Group 7: the H$_2$ emission.} The H$_2$ emission is located furthest into the PDR.\\

We also investigated the profiles of the PAH bands and 15-18 \mum\, PAH plateau. All PAH bands have a symmetric band profile except for the 16.4 \mum\, band, which is slightly asymmetric. We did not observe any changes in the band profiles spatially across the maps. Likewise, we found that the profile of the 15-18 \mum\, PAH plateau does not vary.  We concluded that the carrier of the underlying 15-18 \mum\, plateau is distinct from the individual PAH bands located on top of it.

\acknowledgements EP owes her genuine gratitude to Martin Houde for his help and support throughout this project. We would like to thank the referee Dr. K. Sellgren whose comments have helped to improve the paper. We very gratefully acknowledge sustained support from the NASA Spitzer Space Telescope General Observer Program. Studies of interstellar PAHs at Leiden Observatory are supported through advanced-ERC grant 246976 from the European Research Council.  LJA gratefully acknowledges sustained support from NASA's Laboratory Astrophysics and Astrobiology Programs.

\end{document}